\documentclass[final,5p,times,twocolumn]{elsarticle}

\usepackage{amssymb}

\usepackage{lineno}
\usepackage{enumitem} 
\usepackage{url}
\usepackage{graphicx}
\usepackage{subfig}
\usepackage{amsmath}
\usepackage{booktabs}
\usepackage[usenames,dvipsnames]{color}
\usepackage[section]{placeins}
\usepackage{hyperref}

\journal{Nuclear Instruments and Methods A}

\begin{document}

\begin{frontmatter}



\title{Performance of an HRPPD in Tesla-scale magnetic fields}


\author[BNL]{B. Azmoun}
\author[JLab]{Y. Ilieva}
\author[BNL]{Y. Jin\corref{cor1}}
\ead{yjin1@bnl.gov}
\author[BNL]{J. Kim}
\author[BNL]{A. Kiselev}
\author[Incom]{A. Lyashenko}
\author[BNL]{B.S. Page}
\author[Incom]{M. Popecki}
\author[BNL]{M.L. Purschke}
\author[Yale]{A. Tamis}
\author[BNLSMD]{V. Teotia}
\author[BNL]{C.P. Wong}
\author[BNL]{C. Woody}

\cortext[cor1]{Corresponding authors}

\address[BNL]{Physics Department, Brookhaven National Laboratory, Upton, NY 11973, USA.}

\address[BNLSMD]{Superconducting Magnet Division, Brookhaven National Laboratory, Upton, NY 11973, USA.}

\address[JLab]{Thomas Jefferson National Accelerator Facility, Newport News, VA 23606, USA.}

\address[Incom]{Incom, Inc., Charlton, MA 01507, USA.}

\address[Yale]{Dept. of Physics, Yale University, New Haven, CT 06511, USA.}

\begin{abstract}
    \noindent   High-Rate Picosecond Photodetectors (HRPPDs) are state-of-the-art microchannel-plate (MCP) photodetectors that offer excellent timing and spatial resolution, together with high single-photon detection efficiency. They are currently being considered for the ePIC experiment at the future Electron Ion Collider (EIC) at Brookhaven National Laboratory.  A key requirement for the application of this technology is reliable operation in a strong magnetic field up to 1.5~T, with magnetic flux lines at an inclination of $\le15^\circ$ to the normal of the MCP surface. Magnetic field-induced distortions of the collected charge in MCP-based detectors can generally be compensated by tuning the operating parameters; however, the objective of this study is to quantify this performance in the case of the EIC-HRPPD, a particular MCP stack-up specialized for operation within ePIC. This photosensor employs a high quantum efficiency photocathode, 10~\textmu m capillary pores, narrow transfer gaps, and a pixelated DC-coupled readout. This article explores the optimal operating parameters (mainly the voltages applied across the gaps and the MCPs) for single photon detection at various inclination angles in a uniform field up to 1.8~T. Ultimately, it was found that the gain and single photon detection efficiency of the HRPPD can be recovered over a range of polar inclination angles up to $\pm35^\circ$. 
\end{abstract}


\end{frontmatter}



\section{Introduction}
\label{sec:intro}

Modern microchannel plate (MCP) based photosensors offer distinct advantages that make them attractive options for radiation detection in high-energy particle physics experiments. The High Rate Picosecond Photon Detector (HRPPD)~\cite{Lyashenko, Jin} is the latest generation MCP-based photosensor from Incom, Inc.~\cite{Incom}, derived from its progenitor, the LAPPD \cite{LAPPD1, LAPPD2, LAPPD3, LAPPD4}. The HRPPD features a large active area (104~mm~$\times$~104~mm) and a low profile (13.4~mm thick), up to $10^7$ gain (enabling single photon detection), sub-millimeter position resolution, and a timing resolution approaching a few tens of picoseconds. The photosensor, or “tile”, incorporates a high quantum efficiency transmissive photocathode, followed by two microchannel plate gain stages, and a charge collection anode (see Fig.~\ref{fig:HRPPD-WorkingPrinciple}). The ability to operate in a strong magnetic field, essential to many such experiments, is also a key feature.  In this article, we explore the performance of a sample HRPPD tile in a uniform static magnetic field, up to 1.8~T.  

Though similar MCP-based detectors (including MCP-PMTs \cite{Burle2004, Burle2005, Lehmann} and LAPPDs) have already been shown to operate well in high magnetic fields \cite{B-Field_MCPPMTa, B-Field_MCPPMTb, B-Field_MCPPMT1, B-Field_MCPPMT2, B-Field_LAPPD}, we report on the particular characteristics of a so-called “EIC-HRPPD”, designed for use in the ePIC experiment at the future Electron Ion Collider (EIC)~\cite{EIC} at Brookhaven National Laboratory. The EIC-HRPPD is designed to operate in the 1.5~T magnetic field of the ePIC solenoid magnet with inclination angles up to $\pm15^\circ$ with respect to the normal of the MCP surface. The introduction of such a strong field distorts the flow of charge through the sensor stack, warranting a new study to both verify the expected functionality and to learn more about the detailed behavior of this specific HRPPD configuration.

As the magnetic field strength increases and the inclination angle becomes larger, the transport of primary photoelectrons and subsequent avalanche electrons from the photocathode to the charge collection anode is progressively modified. The trajectory of each electron is influenced by helical gyration, parallel acceleration, and $ E \times B$ drift, resulting in significant reductions in gain, broadening of the transit time spread, and a lateral displacement of the charge cloud with a reduced footprint on the anode plane. As each electron undergoes tighter helical-like motion, the incident angle at which it strikes the functionalized surface of the MCP capillary can deviate substantially from the optimized angle at zero-field, resulting in a reduced probability for releasing secondary electrons\cite{Angle1}. Furthermore, since the axis of the helical motion is aligned with the direction of the magnetic field, the electron trajectory may be partially or fully aligned along the capillary axis, resulting in fewer interactions with the capillary surface and the release of fewer secondaries. Moreover, this helical motion has a focusing effect on the generation of the charge cloud by constraining the transverse component of the electron recoil scattering angle.

The EIC-HRPPD tile was customized to mitigate such distortions and make the device more resilient to strong magnetic fields. The design includes the reduction of the MCP pore diameter compared to earlier sensor designs like that of the LAPPD, which increases the likelihood of an electron interaction with the functionalized layer of the capillary in the presence of a high field to maintain the gain. Smaller pores also restrict ion backflow, which helps to protect the photocathode from ion bombardment and extends its lifetime, while imposing a reduced ballistic deviation for the secondary electrons, which improves both spatial and timing resolutions. In addition to a functionalized emission layer for enhanced gain, a specialized atomic layer deposition (ALD) treatment of the capillary surface adds a resistive layer that significantly reduces the recharge time compared with untreated surfaces \cite{Popecki2016Microchannel} for improved rate capability \cite{AGARWALA2027171914}. Additionally, by reducing the width of the transfer gaps, the EIC-HRPPD is less susceptible to magnetic deflection, resulting in smaller lateral charge displacement and improved transit time spread. Finally, the DC-coupled pixelated readout in this device generally produces smaller pad cluster sizes compared with a comparable capacitively coupled one, resulting in lower occupancy and less overlap between reconstructed hits (though in strong magnetic fields, the cluster size distribution is dominated mostly by single pad hits) . While there are certainly technical hurdles with employing a moderately dense DC-coupled readout in a vacuum tube device, the choice was ultimately made to avoid the technical challenges particular to AC-coupled readouts. The latter would include the deposition of a highly uniform resistive layer on the internal anode plate for effective charge evacuation, and the demand to prevent charge pile-up and baseline shifts in high-rate environments.

As such, HRPPD tiles are the baseline sensor modules for the ePIC pfRICH detector \cite{pfRICH} and are currently under consideration for the ePIC DIRC \cite{DIRC}, and are candidates in other RICH applications, such as the upgrade option for Belle II ARICH at the SuperKEKB accelerator \cite{belleII}. Since each detector system has unique and varied requirements for the photosensor, the goal of this study is to quantify the performance of the EIC-HRPPD over a broad range of operating parameters, mainly a span of voltages applied across the MCPs and transfer gaps, and as a function of the magnetic field strength and tile orientation. In doing so, we attempt to identify a stable operating point at each field value and inclination angle, characterized by adequate gain and photon detection efficiency, and optimal timing resolution. It should be noted that the beam spot diameter of our laser system is too small to allow for adequate charge sharing between adjacent read-out pixels; therefore, this test is not amenable to studies of position resolution under realistic experimental conditions.

\section{Experimental Apparatus}

\subsection{HRPPD Sensor}
\begin{figure}[ht]
    \centering
    \includegraphics[width=\linewidth]{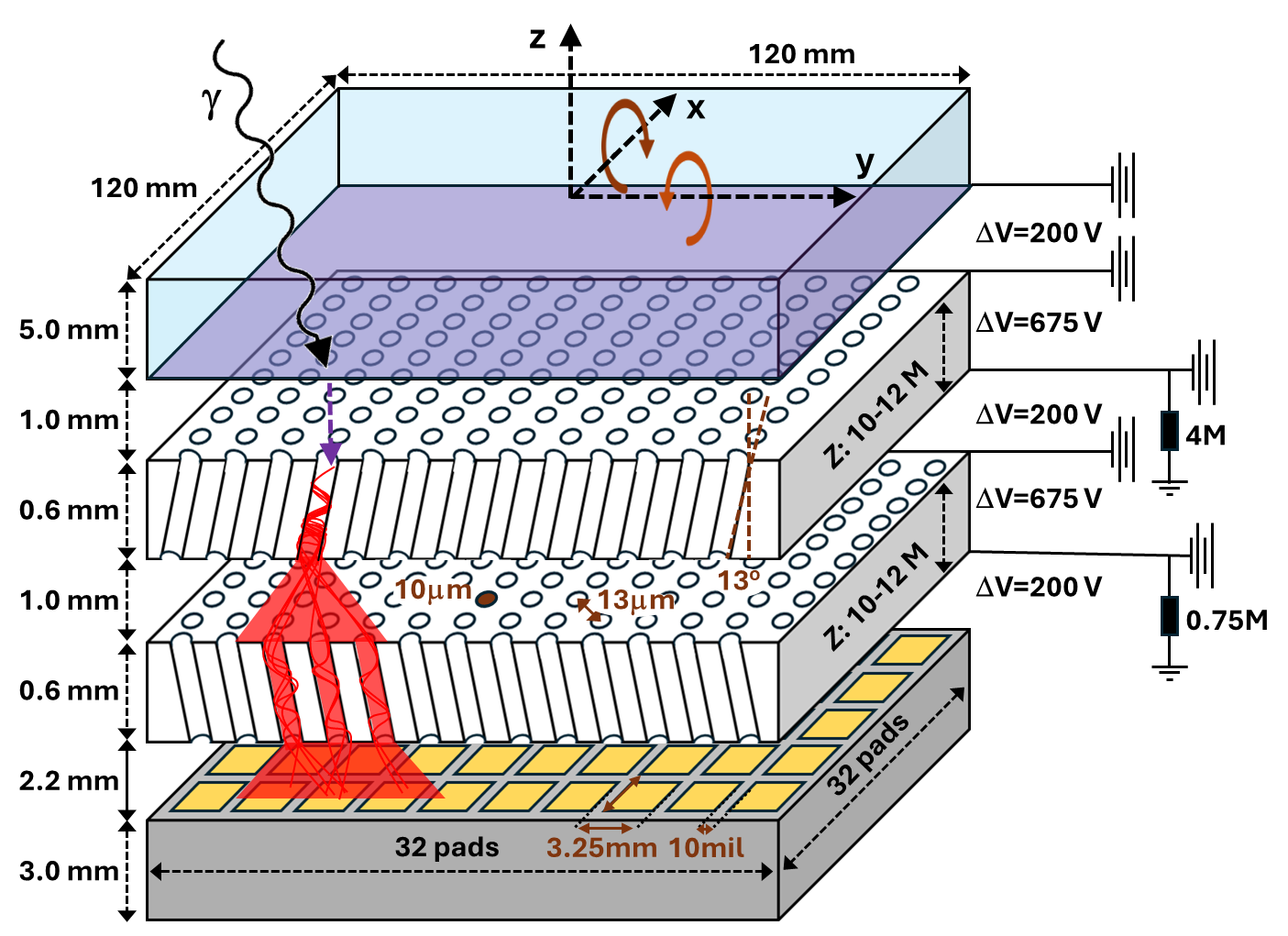}
    \caption{Sketch of the HRPPD inner structure (not to scale). The stackup from top to bottom consists of a fused silica window, followed by two MCP layers and the anode plate. The circular arrows represent rotations about the X- and Y-axes. The dimensions shown for the window represent the physical size of HRPPD, and all dimensions are nominal and accurate to within 10\%. The HV configuration is shown on the right with typical values of the potentials applied across the MCPs and all gaps. (200–675–200–675–200 from bottom to top.)}
    \label{fig:HRPPD-WorkingPrinciple}
\end{figure}

\begin{figure}[ht]
    \centering
    \includegraphics[width=\linewidth]{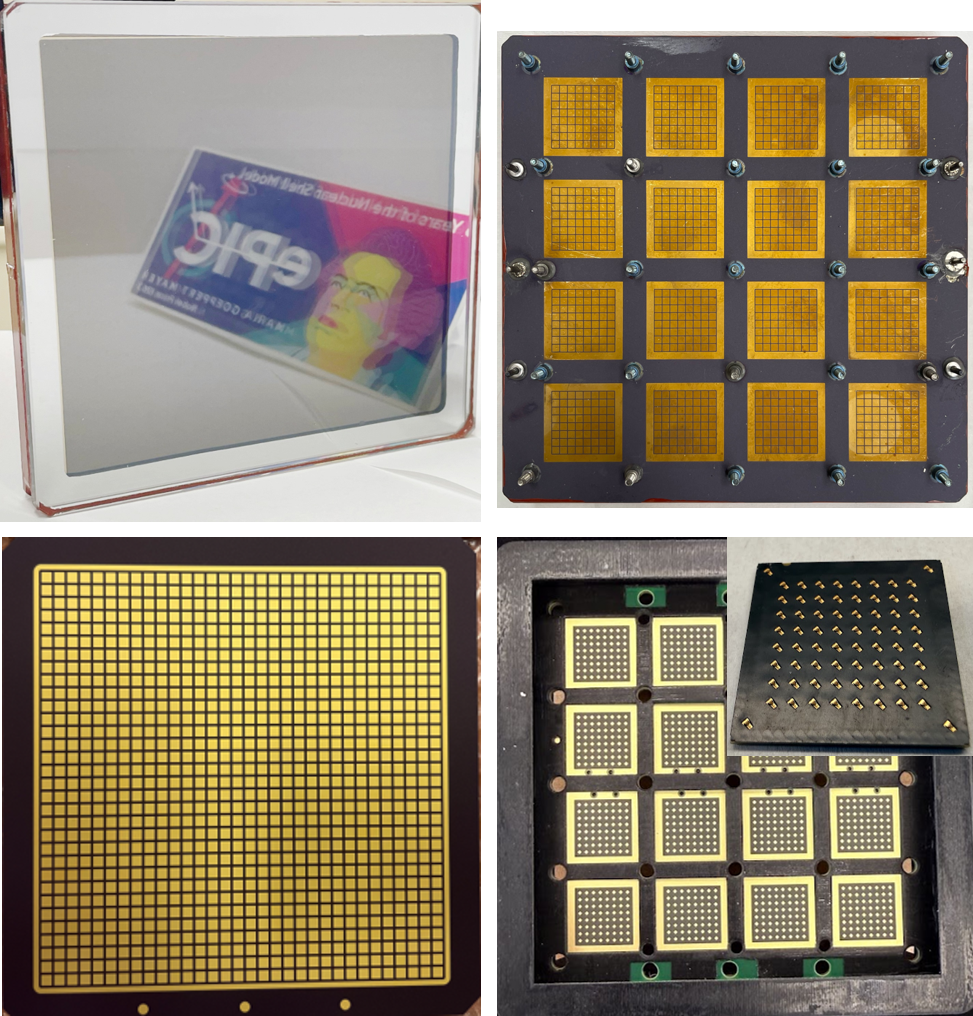}
    \caption{Top-left: HRPPD fused silica entrance window; Bottom-left: pixelated interior surface of ceramic readout plate; Top-right: external side of readout plate with groups of $8 \times 8$ DC-coupled signal contacts; Bottom-right: intermediate readout PCB used to interface contact pads with MCX signal cables via compression interposer, which acts as a signal bridge (inset image).}
    \label{fig:PlaceHolder__HRPPD-Tile_roPCB}
\end{figure}

The EIC-HRPPD sensor tile used in this study was selected from a test batch of seven HRPPDs produced for the EIC collaboration in 2024. Relevant features of these tiles include narrow pore (10~\textmu m) -- small pitch (13~\textmu m) MCPs, thin inter-electrode gaps (1--2~mm), and a chevron-style MCP configuration, with a tilt angle of $13^\circ$, as shown in Fig.~\ref{fig:HRPPD-WorkingPrinciple}. The tile features a semi-transparent photocathode material (multi-alkali, $Na_{2}KSb$) deposited on the inside surface of a fused silica window, which also provides the vacuum seal for one side of the tile. Adjacent to the interior window surface are two MCP gain stages, followed by a high temperature co-fired ceramic (HTCC) readout plate that serves as the bottom vacuum seal. The internal plate surface provides the substrate for the metalized field termination plane, consisting of an array of $32 \times 32$~square charge collection pads with a pitch of 3.25~mm. Each pad is DC-coupled to a contact electrode on the external surface via an embedded routing layer (optimized for minimal cross-talk and impedance matching), which passes the signal via a compression interposer to an array of matching contacts on an intermediate PCB. This PCB then routes the signals to an MCX patch panel board connected to a set of coaxial cables that deliver the raw charge to the front-end electronics. Photos of the physical tile and readout board are shown in Fig.~\ref{fig:PlaceHolder__HRPPD-Tile_roPCB}. Key performance characteristics of this device include high gain ($\sim5\times10^6$), high quantum efficiency ($>30\%$ at 365~nm), excellent single photoelectron timing resolution (15--20~ps), sub-mm position resolution, and exceptionally low dark count and afterpulsing rates ($< 1.8$~kHz/cm\textsuperscript{2} and $<2.8\%$, respectively) \cite{Lyashenko}. 

\FloatBarrier

\FloatBarrier

\subsection{Experimental Method}
\begin{figure}[ht]
    \centering
    \includegraphics[width=\linewidth]{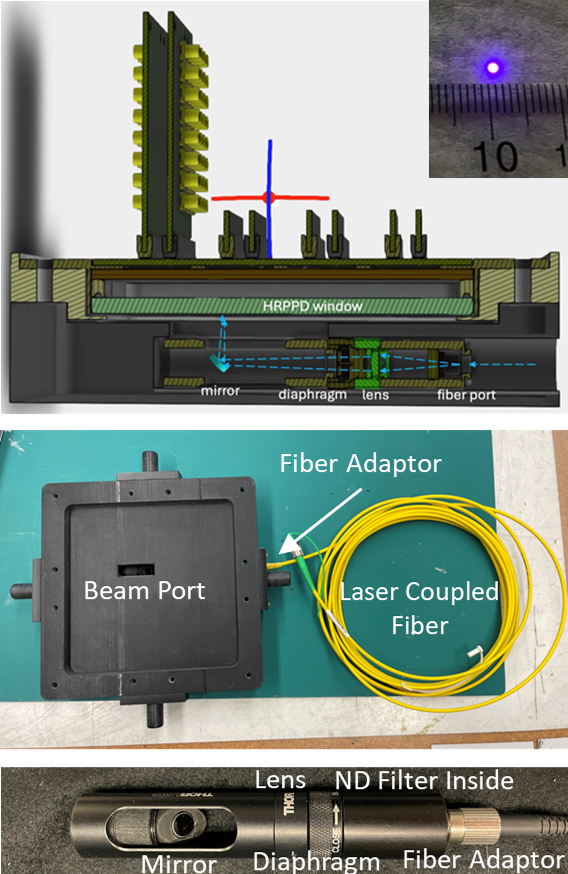}
    \caption{Images of the optical setup. Top: the cross-sectional view of the HRPPD enclosure reveals the optical path of the laser beam delivered by a single-mode fiber. Middle and Bottom: The fiber aperture is re-imaged near the photocathode surface using a focusing lens and a mirror to steer the beam by $90^\circ$ through the beam port and into the HRPPD enclosure. An image of the beam spot is seen at the top right with a diameter of about 1.5~mm.}
    \label{fig:Optical Setup}
\end{figure}

The basic scheme for quantifying the performance of the HRPPD sensor in the magnetic field consists of establishing a laser beam spot on the photocathode surface, then evaluating the sensor's response to the ensuing photoelectrons. The beam spot is approximately 1.5~mm in diameter and is delivered to the photocathode by a single-mode fiber-coupled laser (described in more detail in Section~\ref{sec: daq and instrumentation} and Section~\ref{sec:time res}). The cone of illumination from the fiber first passes through a focusing lens and is then reflected by a $45^\circ$ mirror, which effectively places a source image of the fiber exit aperture very close to the photocathode surface, as shown in Fig.~\ref{fig:Optical Setup}. The laser beam is tuned to produce single photon pulses for the purpose of quantifying the single phototelectron (SPE) response of this setup. Measured pulse height distributions of the total collected charge verify operation in the SPE regime and are shown in Section~\ref{sec:gain}. These are described by Poisson-distributed primary-photoelectron statistics with a mean less than 0.1.    

\FloatBarrier

\subsection{Test Setup}
The HRPPD test setup was placed in a magnetic field with a uniformity of $O(10^{-3})$, while simultaneously acquiring data with the tile in a given rotational orientation, as defined in Fig.~\ref{fig:HRPPD-WorkingPrinciple}, and with a given high voltage configuration, as described in Section~\ref{sec:gain}. 
The measurements were carried out at the BNL Superconducting Magnet Division (SMD) and employed a warm 18D72 type dipole magnet capable of providing approximately a 2 T field in the center of the bore. The distance between pole tips was only 6 inches, which necessitated a test setup as compact as possible and limited the maximum inclination angles to about $30^\circ$--$35^\circ$. 

As shown in the photos in Fig.~\ref{fig:SMD-Collage}, the HRPPD test setup was staged onto a pair of sliding rails, which allowed one to manually fix the inclination angle of the setup, then slide the setup into the magnet bore and begin a data run. This process was iterated systematically over a matrix of field strength and voltage configuration ranges, as described in detail in Section~\ref{sec: measurements}. The inclination angle of the HRPPD tile was set relative to the magnetic field direction through a rotation around either the X-axis or the Y-axis, as shown in Fig.~\ref{fig:HRPPD-WorkingPrinciple}. In both cases, the angle was set by placing a pin into one of a series of holes spanning a metal arc connected to the black HRPPD enclosure, at discrete intervals of $2.5^\circ$ (shown in the middle-right image in Fig.~\ref{fig:SMD-Collage}).

\subsection{DAQ and Instrumentation}\label{sec: daq and instrumentation}

The instrumentation employed in this study mainly consists of a data acquisition system, a high voltage system, a fiber-coupled laser, a 3-axis Hall probe, and a magnet power supply delivering high current to the magnet coils. A versatile and lightweight data acquisition (DAQ) software program is employed to orchestrate much of this instrumentation to produce the data files used for both online and offline analysis.

The magnet power supply is part of legacy equipment at SMD, retrofitted with a modern graphical user interface for adjusting and monitoring the current (\textit{i.e.}, up to 2000 A at maximum field). The applied magnet current is the underlying independent variable responsible for the magnetic field strength in the series of measurements presented in this work. However, we rely on the Hall probe to provide a direct measurement of the field strength, which also accounts for any uncertainty related to magnetic hysteresis.

\begin{figure}[ht]
    \centering
    \includegraphics[width=\linewidth]{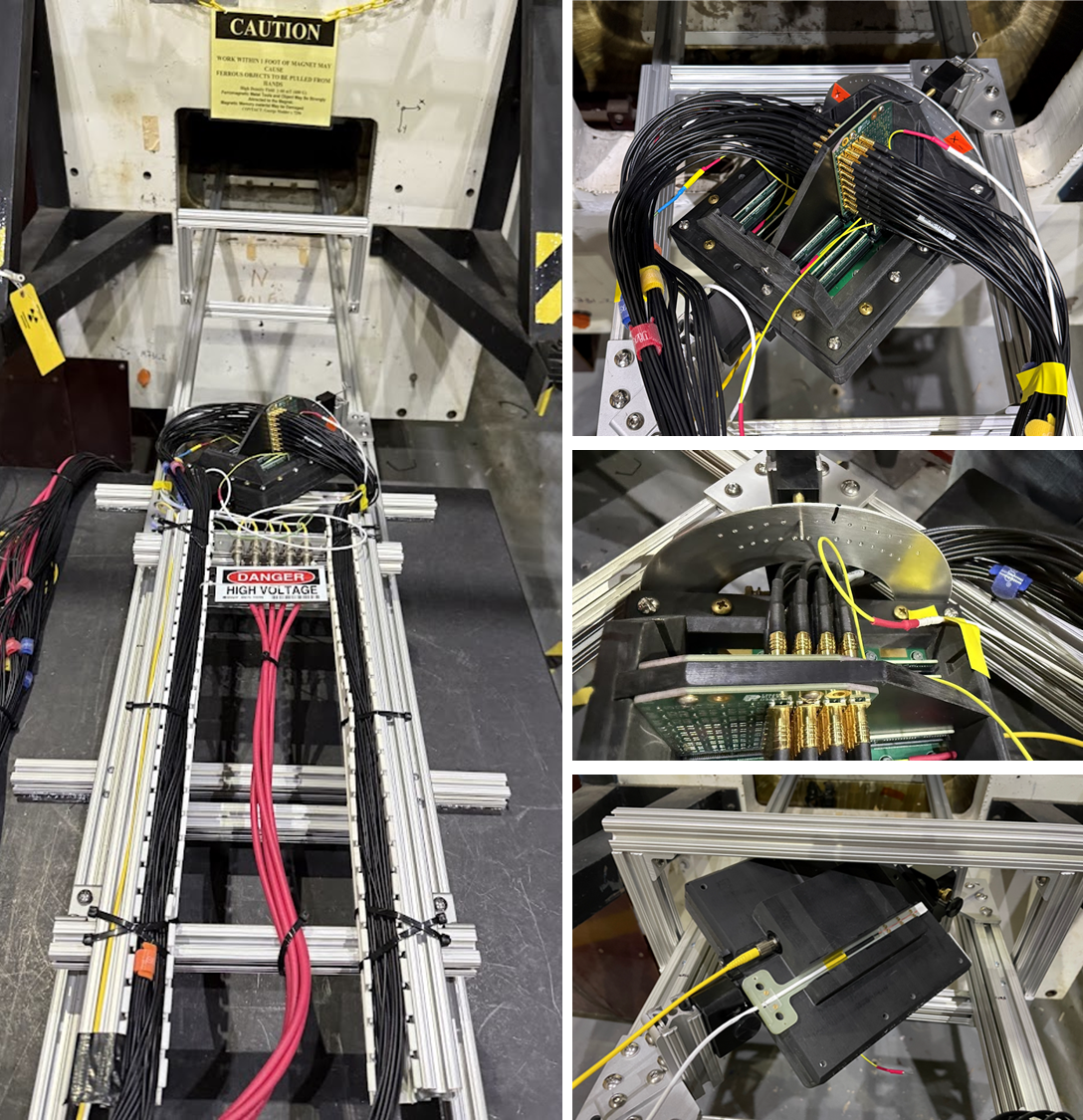}
    \caption{Experimental apparatus at the BNL SMD. The photo on the left shows the HRPPD enclosure (shown zoomed in on top-right) sitting atop a rail system used to insert and retract the enclosure from the magnet bore. The middle-right and bottom-right photos depict the pinholes used to set the inclination angle and a zoom-in on the Hall probe mounted to the bottom of the enclosure, respectively.}
    \label{fig:SMD-Collage}
\end{figure}

The Hall probe (Senis, 3MH6 Teslameter) provides a high-precision B-field measurement along three orthogonal axes. The probe itself is secured to a flat, recessed surface of the HRPPD enclosure (as shown in the bottom-right image of Fig.~\ref{fig:SMD-Collage}) to ensure the probe's coordinate system axes are parallel to that of the HRPPD's sense plane. As such, the three components of the field reported by the probe may be used to determine the set inclination angle of the HRPPD relative to the magnetic field direction, to within sub-percent precision, as described further in Section~\ref{sec:inclination angle}. 

The laser system used in this study is a 420~nm picosecond pulsed diode laser (NKT Photonics, PiLas) whose output is coupled to a single-mode fiber. The fiber is 10~ft. long and allows for the laser to be positioned well outside of the magnetic field while delivering photons to the HRPPD photocathode. The intensity of the laser beam can be tuned with the laser controller in order to operate in the SPE regime (see Section \ref{sec: measurements}). It must be noted that the single photon timing jitter measured with this laser is dominated by its relatively broad pulse width of 35~ps full width at half maximum (FWHM). This imposes severe limitations on the timing measurements, as detailed below in Section \ref{sec:time res}.

The HV system (CAEN, R8034DN) independently energizes each of the five electrodes in the HRPPD stackup to produce the typical potentials shown in Fig.~\ref{fig:HRPPD-WorkingPrinciple}. Each channel is trip protected and can deliver up to 1 mA. The HV connection to the bottom electrode of each MCP (as shown in the same figure) includes a bleeder resistor to ground to compensate for the fact that two HV supply outputs are coupled across the resistive load of each MCP. Since each HV channel cannot sink current, the bleeder resistors provide an alternate path to ground. This allows for the possibility to independently set the HV values of each MCP electrode and avoid this limitation.  

The DAQ hardware system mainly consists of a DRS4-based waveform digitizer module (CAEN V1742). The module plugs into a standard VME crate and accommodates 32~input channels, each of which is connected to a subset of 1,024~pads on the HRPPD sense plane using 10~ft. coaxial cables. Each cable connects to a MCX patch panel board, which routes signals from the HRPPD, as described earlier. A pulse generator is used to send triggers to both the laser and the digitizer in parallel. The 32~analog pad signals are then simultaneously digitized across 1,024~samples at a sampling rate of 5~GHz. The corresponding ADC data are subsequently saved to disk at a rate of about 1.5~kHz. One typical data file contains 10$^5$ events. The pedestal fluctuation is measured to be 0.35~mV, in agreement with DRS4 specifications~\cite{DRS4noise}.

The data-taking software consists of a data acquisition program, RCDAQ \cite{RCDAQ} and a data quality monitoring program based on the PMONITOR library \cite{Pmonitor}. The recorded data are written to disk and processed offline with custom scripts using the CERN ROOT analysis framework. RCDAQ provides data streams for online monitoring purposes that can feed data into the PMONITOR package. This allows for real-time monitoring of the data.

\FloatBarrier

\section{Measurements and Results}\label{sec: measurements}

\subsection{Inclination Angle Determination}
\label{sec:inclination angle}

As described above, a 3D Hall probe was attached to the HRPPD enclosure, enabling a precise measurement of the magnetic field's magnitude and direction. At each magnet configuration, a deviation of less than 0.3\% from the average field strength was observed over time. At each nominal tilt setting, the $x$, $y$, and $z$ components of the measured magnetic field were used to determine a value of tilt relative to the normal of the MCP surface. On average, this determined value was -1.41~degrees offset from the nominal setting along the $x$-tilt axis and -0.43~degrees along the $y$-tilt axis. The nominal angle settings are established by using the pin-holes described above, where the quoted offsets are the calculated systematic errors associated with the mechanics. All figures are shown with these offsets accounted for.

\subsection{Gain}\label{sec:gain}

\begin{figure}[ht]
    \centering
    \includegraphics[width=\linewidth]{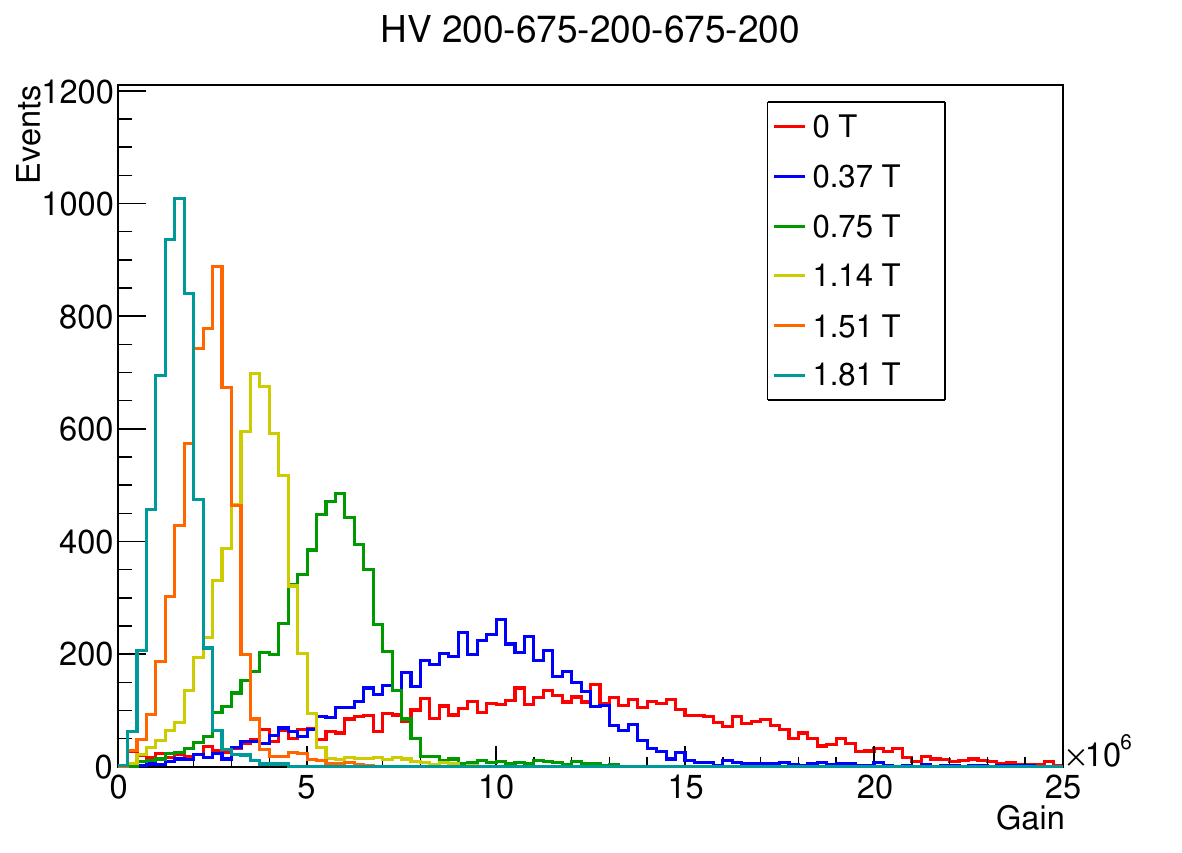}
    \caption{Gain spectra measured for different magnetic field strengths at $0^\circ$. The spectrum of gain (red) acquired in the absence of a magnetic field is horizontally scaled by a factor of 0.3 for visual comparison. In the presence of a magnetic field, the gain spectra exhibit a non-Gaussian shape. Pedestals are completely cut off by the thresholds due to the low noise of the readout system.}
    \label{fig:spectra}
\end{figure}

A peak-searching algorithm is used to identify signal pulses that exceed a predefined threshold and exhibit well-defined falling and rising edges. The threshold is set between 2 and 4~mV for different datasets, depending on the average amplitude of the signal pulses. The single-photoelectron signal amplitude is typically 60 mV for a gain of about 2.5$\times 10^6$. The pulse charge is integrated from the peak position in both directions until the waveform crosses the baseline. A variety of operating conditions were investigated as a function of the gain (pulse charge / elementary charge). The gain spectra obtained under six magnetic field configurations are shown in Fig.~\ref{fig:spectra}, where the values listed in the title, from left to right, correspond to the voltages applied between the anode and the exit surface of the bottom (second) MCP, across the bottom (second) MCP, the transfer gap, top (first) MCP, and between the entry surface of the top (first) MCP and the photocathode, respectively. 

\FloatBarrier

\begin{figure}[ht]
    \centering
    \includegraphics[width=\linewidth]{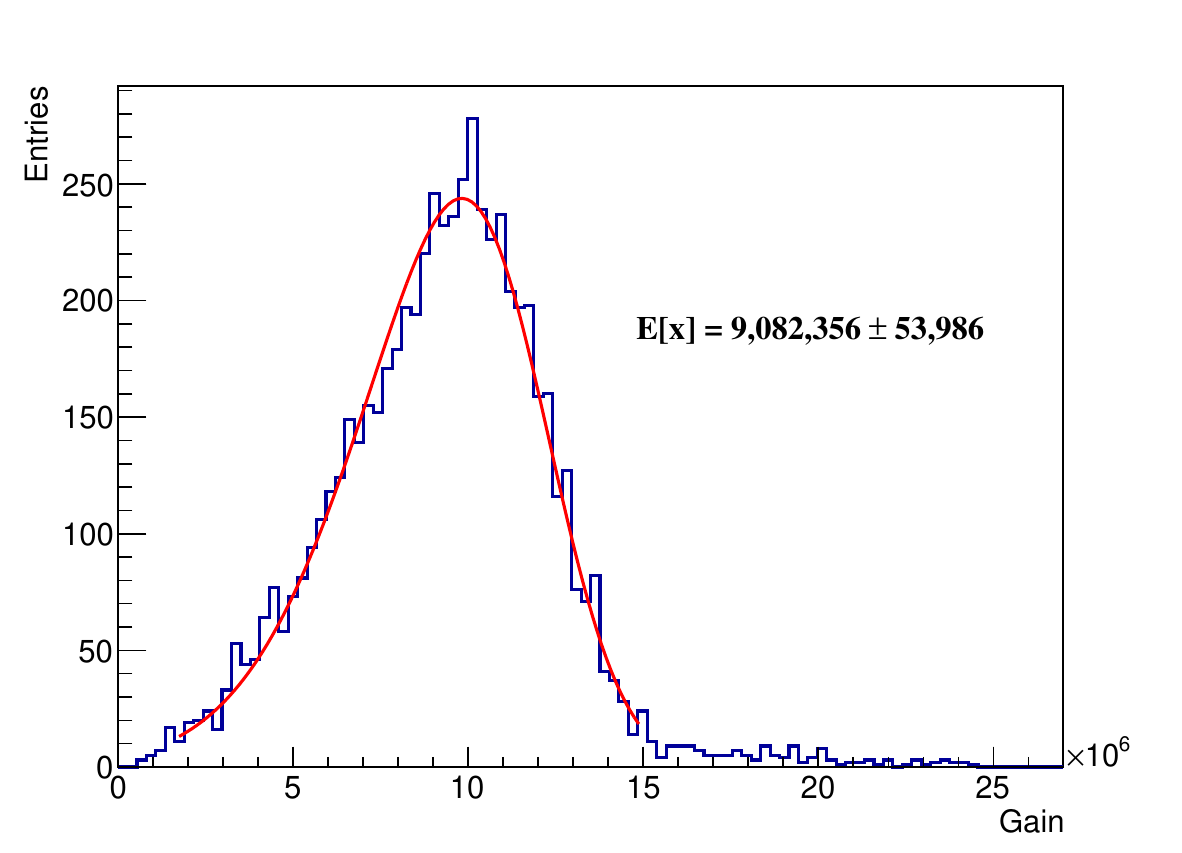}
    \caption{Gain spectrum of a typical run, fitted with a skewed Gaussian. Entries in the right tail are attributed to double-photoelectron events.}
    \label{fig:fit}
\end{figure}

The resulting gain spectrum is fitted with an empirical skewed Gaussian function, as shown in Fig~\ref{fig:fit}:
\begin{equation}
f(x;\xi,\omega,\alpha)
=
\frac{1}{\omega\sqrt{2\pi}}
\exp\left[
-\frac{(x-\xi)^2}{2\omega^2}
\right]
\left[
1+\operatorname{erf}
\left(
\frac{\alpha(x-\xi)}
{\omega\sqrt{2}}
\right)
\right],
\end{equation}
the mean of which is taken as the estimated gain:
\begin{equation}
E[x]
=
\xi
+
\omega
\sqrt{\frac{2}{\pi}}
\frac{\alpha}{\sqrt{1+\alpha^2}}.
\end{equation}

\FloatBarrier

\begin{figure}[ht]
    \centering
    \includegraphics[width=\linewidth]{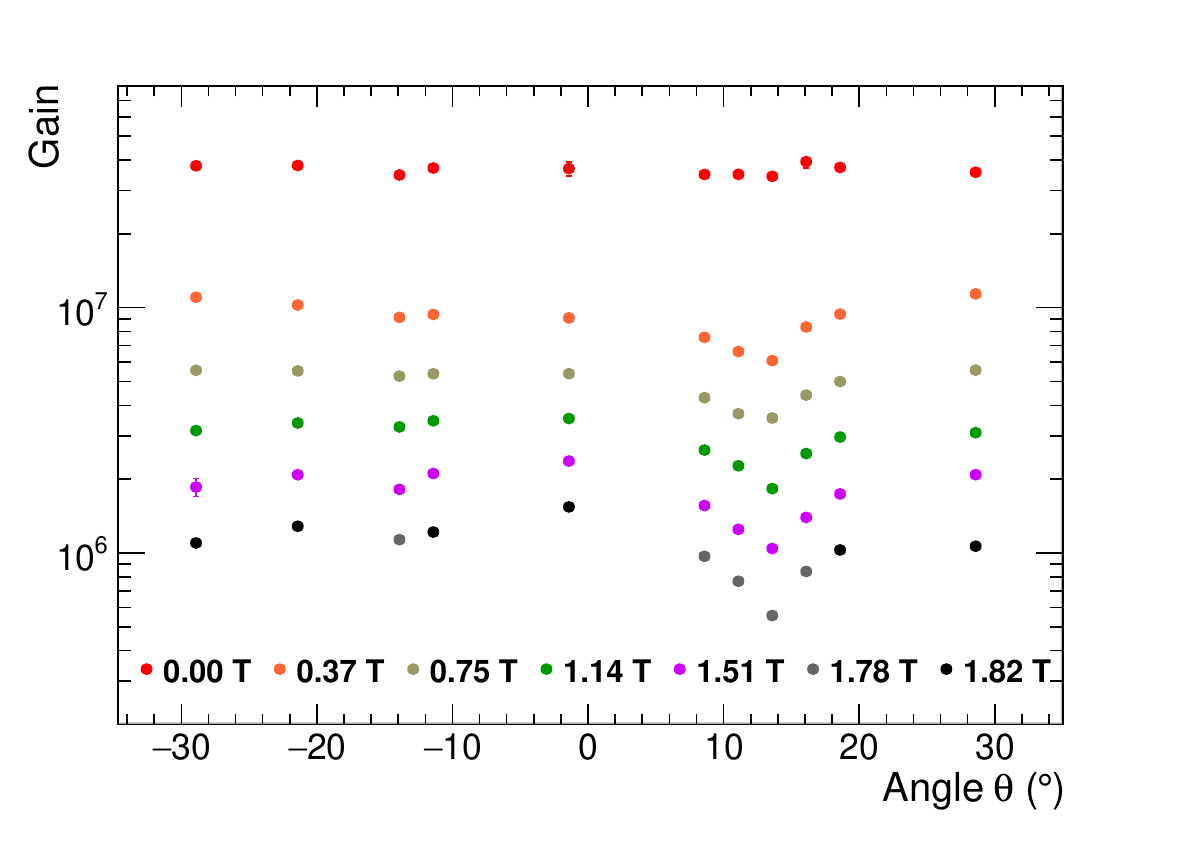}
    \caption{Gain as a function of inclination angle about the X-axis, with a photocathode voltage of $200$~V, and symmetric bias voltages of $675$~V on the MCPs. Note that the error bars represent only the uncertainties from the fit. (Due to the maximum safe operating load of the magnet power supply, the magnetic field was limited to 1.78~T for some angles.)}
    \label{fig:gain_angleX}
\end{figure}

\begin{figure}[ht]
    \centering
    \includegraphics[width=\linewidth]{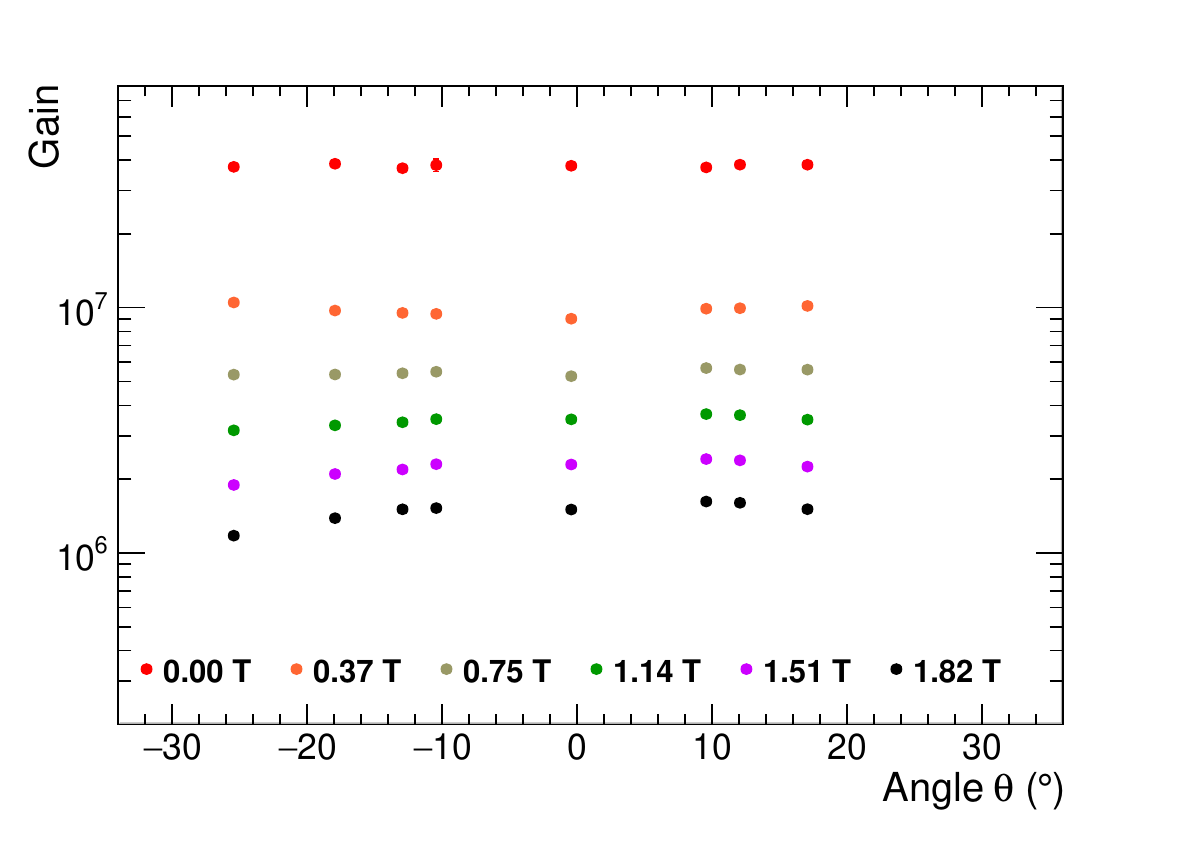}
    \caption{Gain as a function of inclination angle about the Y-axis, with a photocathode voltage of $200$~V, and symmetric bias voltages of $675$~V on the MCPs. Note that the error bars represent only the uncertainties from the fit.}
    \label{fig:gain_angleY}
\end{figure}

Figure~\ref{fig:gain_angleX} shows the HRPPD gain measured as a function of the nominal inclination angle about the X-axis for several magnetic field strengths. A pronounced gain minimum is observed near $13^\circ$, and a minor minimum is seen near $-13^\circ$, corresponding to the alignment of the magnetic field (z direction) with the capillaries of the first and second MCPs, respectively. The reduction in gain at these angles indicates that the alignment of the magnetic field with the MCP capillaries suppresses the number of interactions that generate secondary electrons, thereby reducing the electron multiplication during the avalanche process, as discussed in Section~\ref{sec:intro}. Notably, the gain dip associated with the alignment of the magnetic field with the first MCP capillaries is significantly more pronounced than that associated with the second MCP.
Figure~\ref{fig:gain_angleY} presents an analogous scan of gain as a function of the inclination angle about the Y-axis, which is perpendicular to the MCP capillary direction. No significant gain dip is observed in this case.

\FloatBarrier

\FloatBarrier

\begin{figure}[ht]
    \centering
    \includegraphics[width=\linewidth]{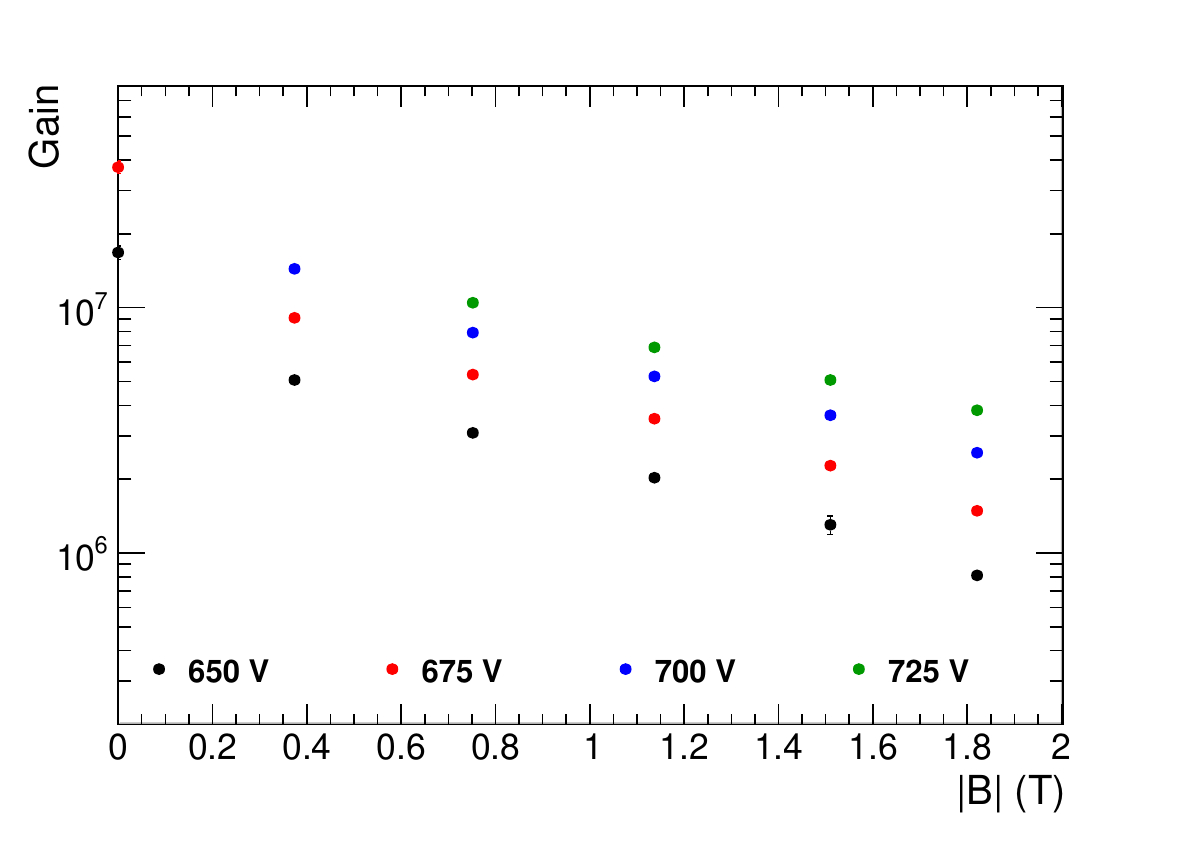}
    \caption{Gain as a function of magnetic field with a photocathode voltage of $200$~V, and symmetric bias voltages on the MCPs. The data shown here were acquired at a nominal inclination angle of $0^\circ$. The same trend is observed over the entire range of inclination angles investigated.}
    \label{fig:gain_bfield}
\end{figure}

\begin{figure}[ht]
    \centering
    \includegraphics[width=\linewidth]{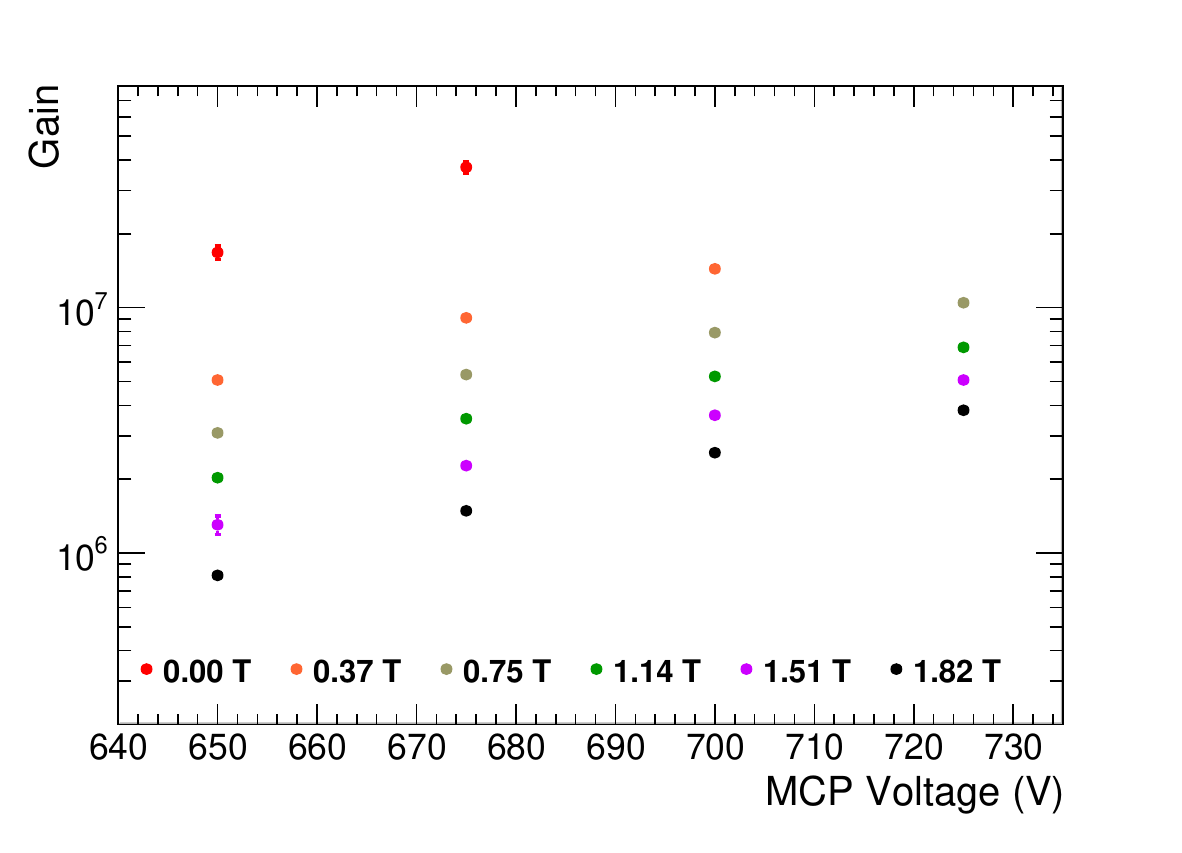}
    \caption{Gain as a function of high voltage across the MCPs with a photocathode voltage of $200$~V.}
    \label{fig:gain_HV}
\end{figure}

To compensate for the gain reduction induced by the magnetic field, as shown in Figs~\ref{fig:gain_angleX} and \ref{fig:gain_angleY}, the bias voltages across the MCPs were increased to restore the signal amplitudes to a required level. Figure~\ref{fig:gain_bfield} demonstrates that an increase of only a few tens of volts is sufficient to recover the nominal operating gain ($>$$10^6$). In particular, a bias voltage of approximately $675$~V across each MCP enables the HRPPD to achieve a gain over $10^6$ in magnetic fields up to $1.8$~T for most angles. The same data are recast as a function of the HV in Fig.~\ref{fig:gain_HV}.

\FloatBarrier

\begin{figure}[ht]
    \centering
    \includegraphics[width=\linewidth]{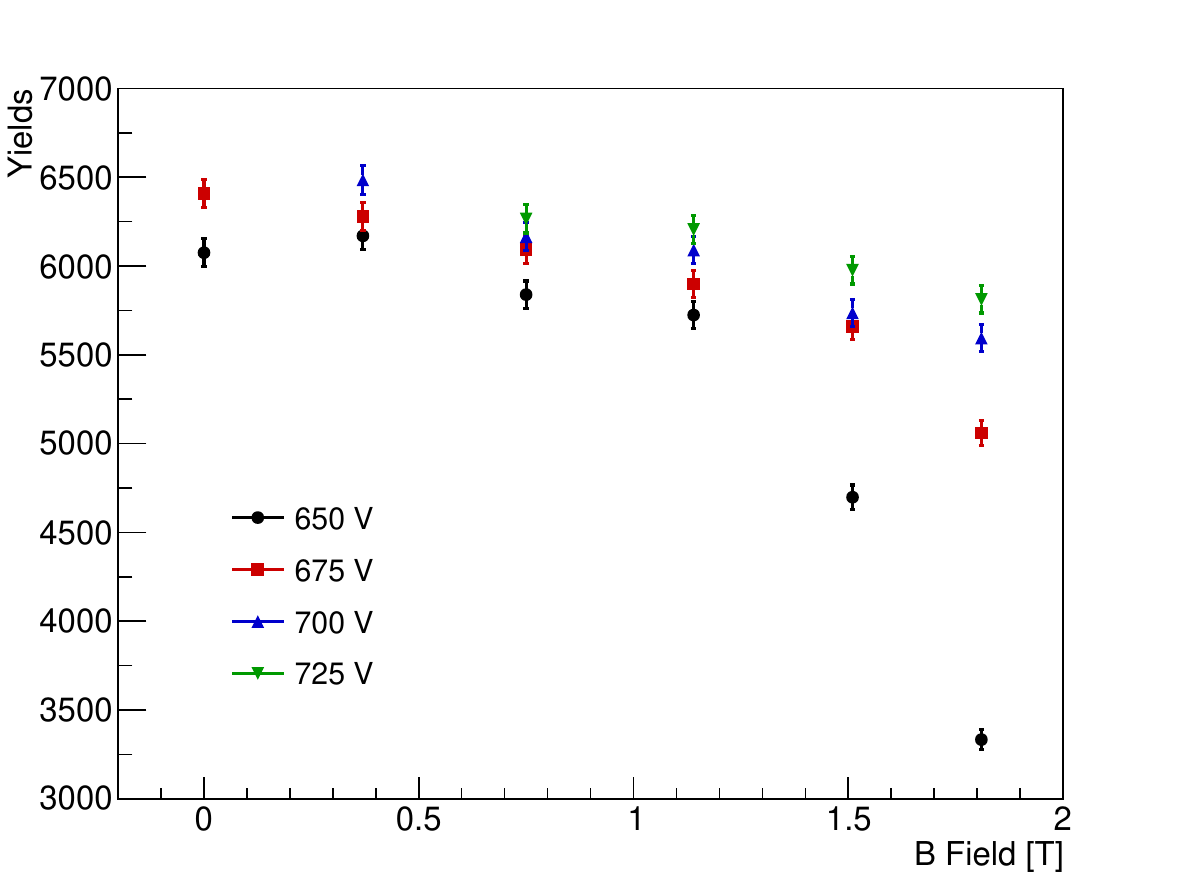}
    \caption{Detected signal yield for different bias voltages across the MCPs and in a range of magnetic fields at $0^\circ$. Note that error bars only indicate statistical uncertainty.}
    \label{fig:detection}
\end{figure}

Similar to the gain, the measured signal yields, determined by directly counting the events of signals exceeding the applied threshold, at different magnetic field strengths are shown in Fig.~\ref{fig:detection}. Increasing the MCP bias voltage effectively compensates for the reduction in signal yield caused by the magnetic field.

\FloatBarrier

\subsection{Afterpulsing rate}

\begin{figure}[ht]
\centering
\includegraphics[width=\linewidth]{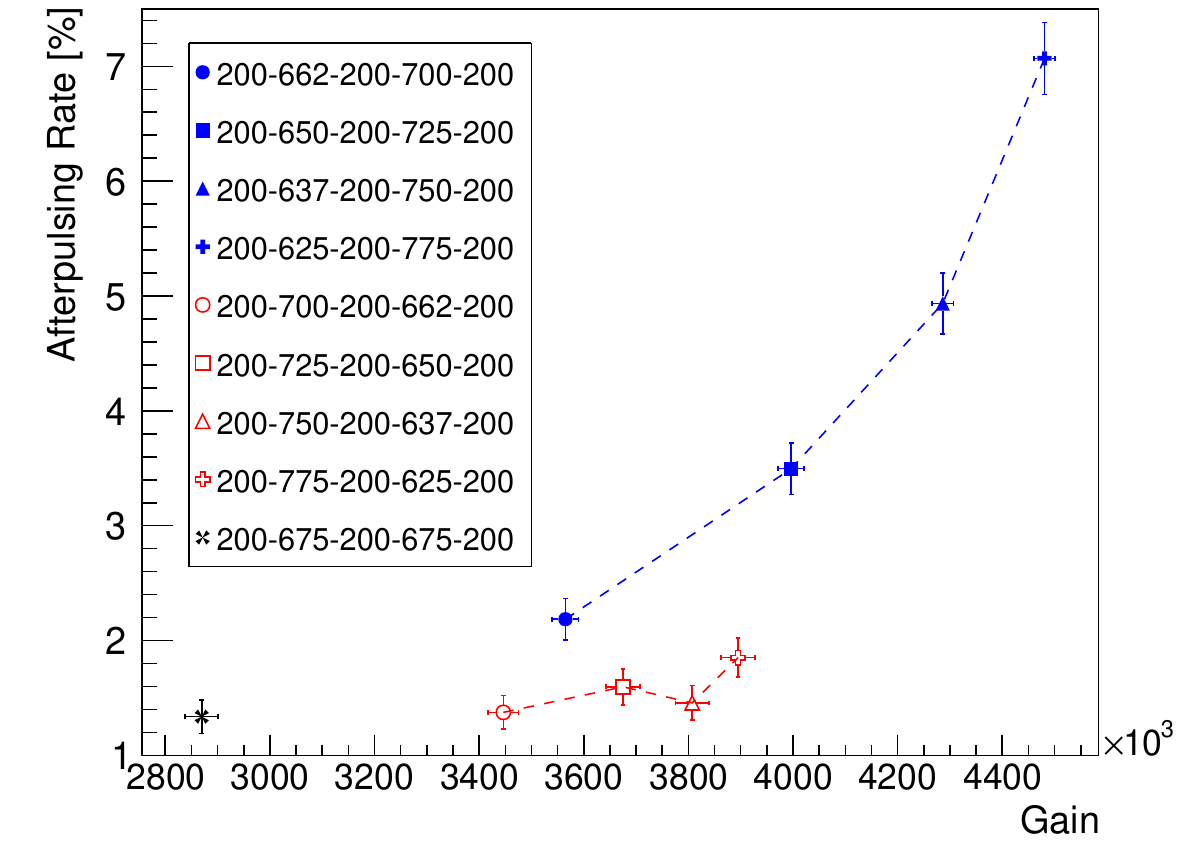}
\caption{Measured afterpulsing rates at 1.3~T for various HV configurations. The four data points connected by the blue dashed line represent HV settings where the first MCP is operated at a higher bias voltage relative to the second MCP. The blue and red data points were obtained using the same selection criteria as those applied to the black data point, which represents the default configuration.}
\label{fig:APR}
\end{figure}

The afterpulsing rate was investigated using the peak-search algorithm within a 70~ns window following the primary signal peak. The amplitude threshold for afterpulse detection was set to be identical to that used for the primary signals. The HRPPD was placed within a uniform magnetic field of 1.3~T. At the nominal operating voltage of 675~V, the afterpulsing rate was measured to be (1.34 $\pm$ 0.15)\%. To identify the primary source of these pulses within the stack, the bias voltage of one MCP was increased in 25~V increments while the voltage on the other MCP was simultaneously decreased in 12~V steps to maintain a consistent order of magnitude for the gain. As illustrated in Fig.~\ref{fig:APR}, the afterpulsing rate exhibits a markedly higher sensitivity to the bias voltage of the first/top MCP (blue) compared to the second/bottom (red), indicating that the first MCP is the dominant contributor to afterpulsing. The weak dependence of the afterpulsing rate on the second MCP bias voltage further indicates that ion feedback from the second MCP is effectively blocked by the first MCP.

\FloatBarrier

\begin{figure}[ht]
\centering
\includegraphics[width=\linewidth]{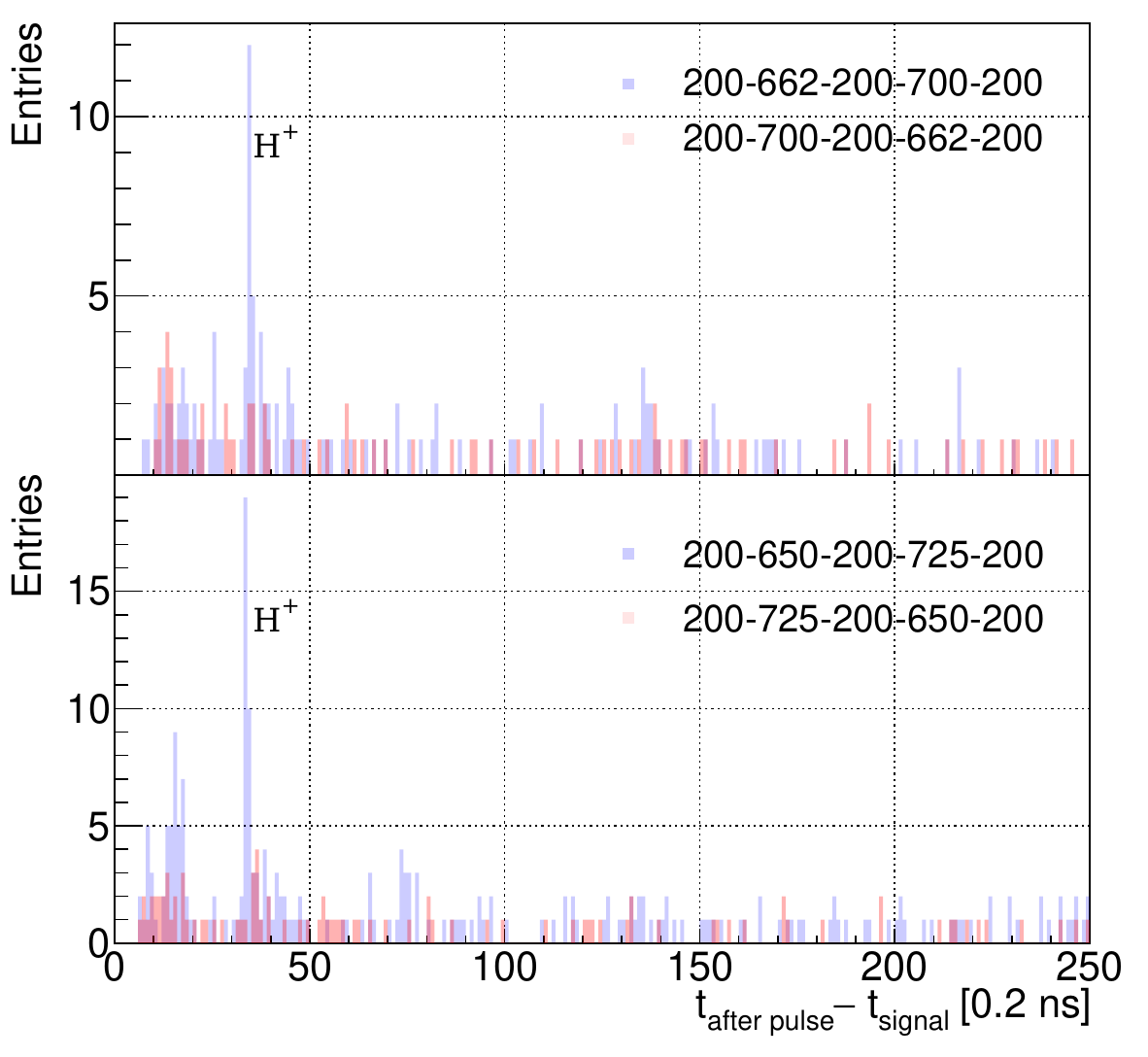}
\caption{Time separation between the primary signal peak and the subsequent afterpulse peak in a 1.3~T magnetic field for four HV settings near the nominal operating HV setting at $0^\circ$.}
\label{fig:TimeSep}
\end{figure}

To further identify the origin of the afterpulses, the time separation between the primary signal peak and the subsequent afterpulse peak was computed. As illustrated in Fig.~\ref{fig:TimeSep}, a distinct peak occurs at approximately 6.4~ns (corresponding to 32 bins in the waveform recorded by the digitizer V1742) when the first MCP is operated at a higher bias voltage. This timing interval is consistent with the flight time of $H^+$ ions originating from the bottom of the first MCP.

\FloatBarrier

\subsection{Dark-count rate} 

\begin{figure}[ht]
\centering
\includegraphics[width=\linewidth]{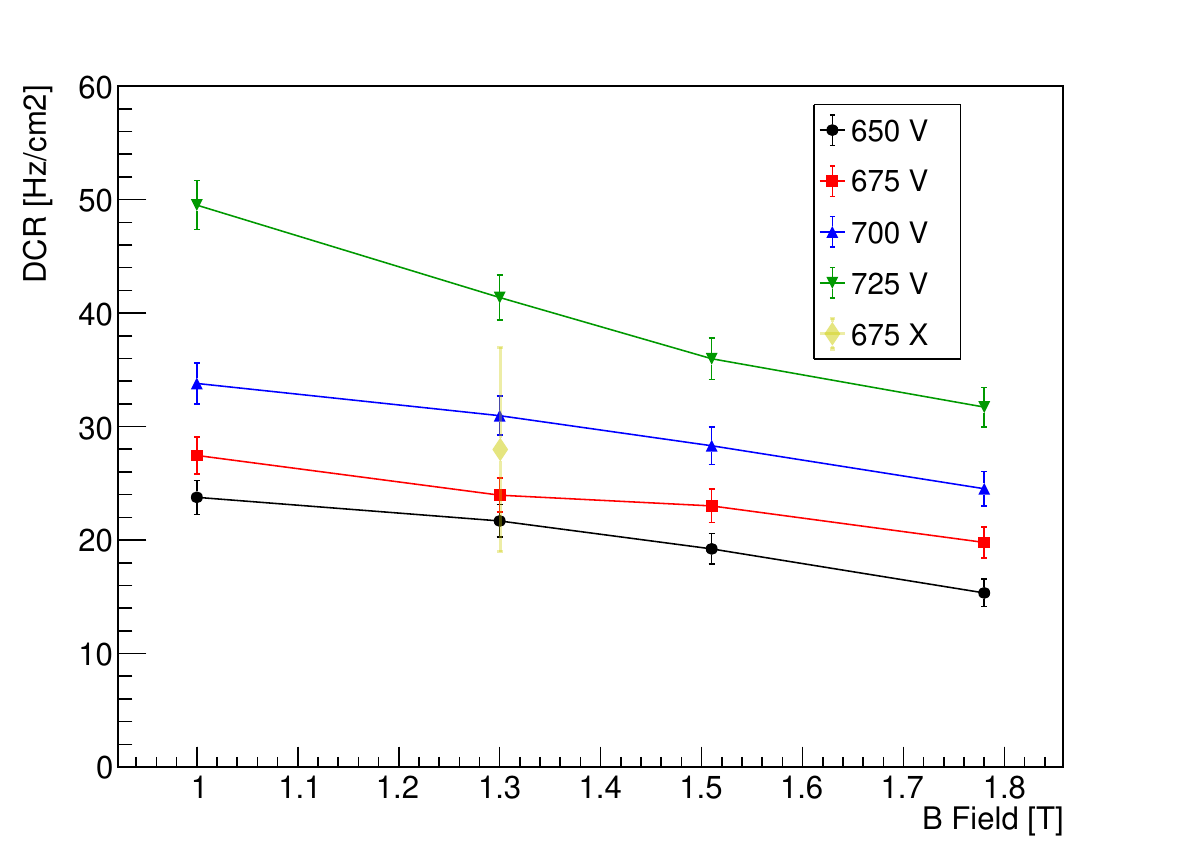}
\caption{Dark count rate of one pixel in a range of magnetic fields at $0^\circ$. Values in the legend indicate the HV across each MCP. The yellow point (675~X) represents the result of the cross-check using the digitizer V1742.}
\label{fig:DCR}
\end{figure}

To estimate the dark-count rate of the HRPPD, a single pixel located away from the laser-illuminated region was connected to an oscilloscope. The oscilloscope was operated in single-trigger mode with a threshold of 4~mV, and the number of pulses recorded over a 100~s interval was used to determine the dark-count rate. Measurements were performed at several magnetic-field strengths, and the results are presented in Fig.~\ref{fig:DCR}. It should be noted that the dark-count rate exhibits spatial non-uniformity across the HRPPD, and therefore the absolute values measured at a single pixel may not be representative of the entire device. Nevertheless, a clear trend is observed: the dark-count rate decreases monotonically with increasing magnetic-field strength and increases monotonically with increasing HV bias, as expected from the behavior of the gain presented above. The observed reduction in the DCR can also be partially attributed to the loss of low-amplitude events due to the applied threshold.
As a cross-check, we also analyzed waveforms recorded by the CAEN V1742 digitizer. A total of 1.1 million events collected in a 1.3~T magnetic field with 675~V applied to each MCP were analyzed, each event containing 32 waveforms corresponding to the 32 readout channels. Using a 100~ns time window preceding the laser pulse, 10 dark-count events were identified, corresponding to a dark-count rate of $28 \pm 9~\mathrm{Hz/cm^2}$, consistent with the result shown in Fig.~\ref{fig:DCR}.

\FloatBarrier

\subsection{Time resolution}\label{sec:time res} 

\subsubsection{Results with Digitized Waveforms} 

\begin{figure}[ht]
\centering
\includegraphics[width=\linewidth]{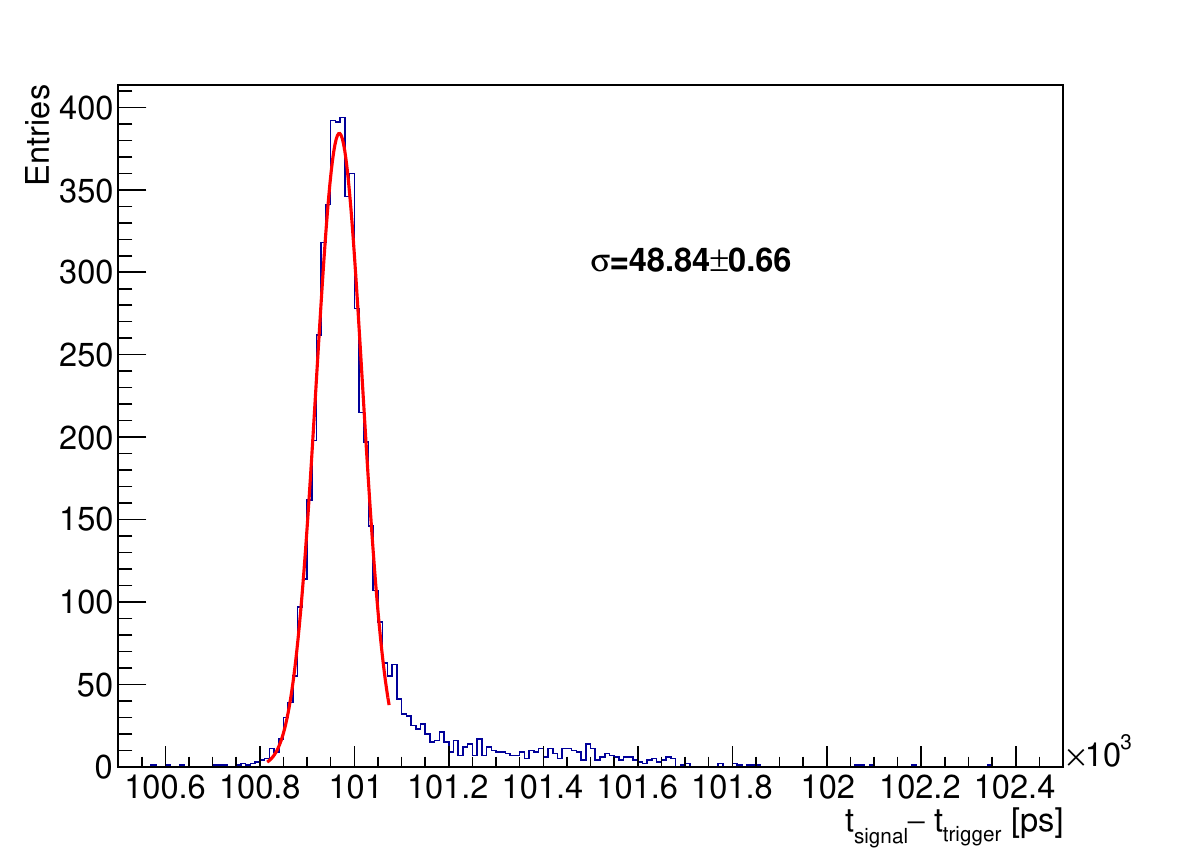}
\caption{Time separation between the trigger and the signal of HRPPD. Entries in the right tail, which are not included in the fitting range, arise from backscattered photoelectrons originating in the interstitial regions of the MCP surface, typically discarded in offline physics analyses.}
\label{fig:Timing1}
\end{figure}

\begin{figure}[ht]
\centering
\includegraphics[width=\linewidth]{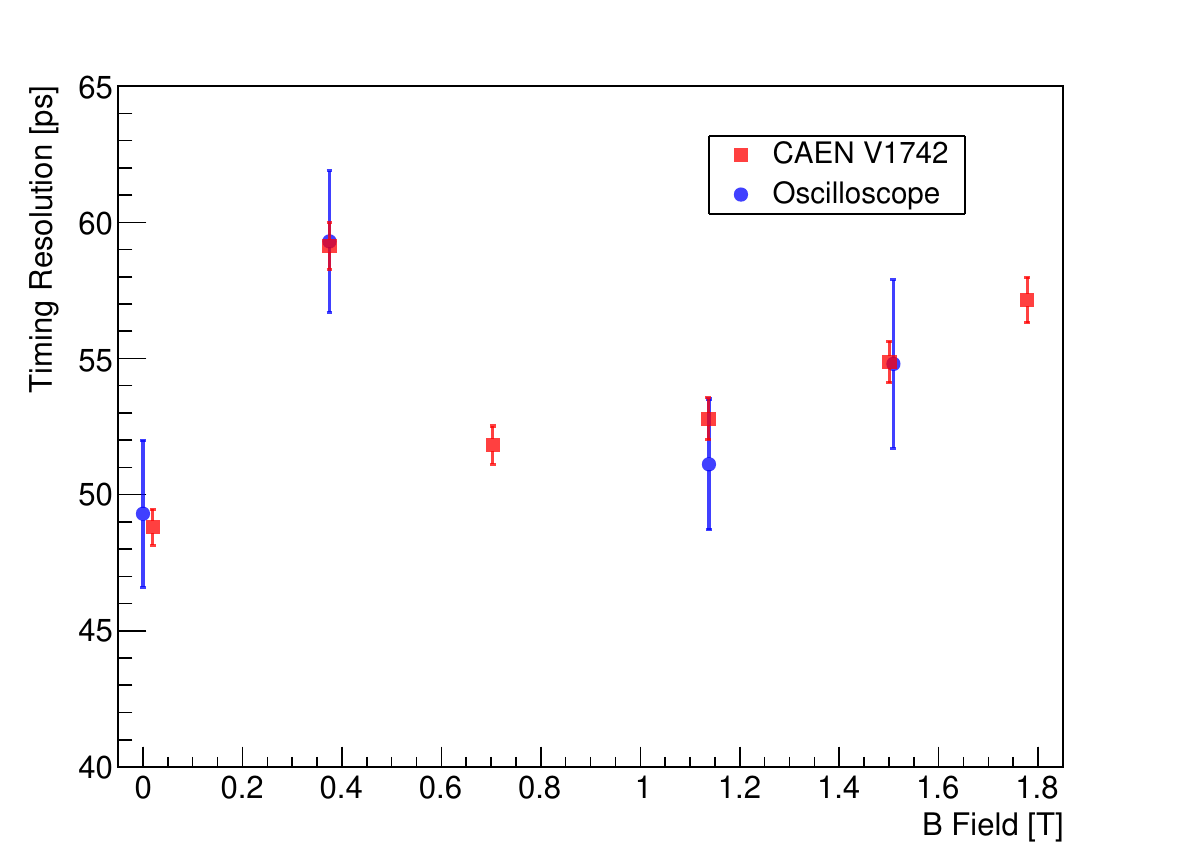}
\caption{Timing resolution across a range of magnetic field strengths at $0^\circ$. Nominal HV setting of 200-675-200-675-200 is used.}
\label{fig:Timing2}
\end{figure}

The CAEN V1742 has 32+2 individual channels, where the first 32 are signal inputs and get connected to the HRPPD pixels. The remaining two inputs can be used to digitize the trigger signals that were sent to the unit. The time interval between the midpoint of the trigger falling edge and the midpoint of the HRPPD signal’s falling edge was calculated, with the variance of this distribution serving as a measure of the timing resolution. To minimize electronic jitter, a standard time correction~\cite{V1742} was applied to the DRS4 chips within the digitizer. Figure~\ref{fig:Timing1} illustrates the time separation for a representative dataset; the distribution is fitted with a Gaussian function, and the resulting standard deviation ($\sigma$) is reported as the timing resolution. The event selection criteria applied here are identical to those used for the gain estimation. The result includes non-negligible timing uncertainty associated with the trigger signal (RMS $\sim$12 ps) and the PiLas pulse width (FWHM $\sim$35 ps), as mentioned in Section~\ref{sec: daq and instrumentation}. Note that the tail of the distribution, which corresponds to backscattered photoelectrons, has been excluded in the fitting range. The timing resolution was also measured across a range of magnetic field strengths, with the results summarized in Fig.~\ref{fig:Timing2}. For future timing characterizations, the implementation of a femtosecond laser is planned to mitigate reducible systematic uncertainty and improve precision.

\FloatBarrier

\subsubsection{Results with Oscilloscope}
In addition to the data taken with the CAEN digitizers, a small dataset was collected with a high-precision Tektronix MSO66B six-channel oscilloscope. The analog waveforms from the NIM trigger signal and the primary illuminated HRPPD pad were sampled by the oscilloscope at a rate of 50 Giga-samples per second (for an absolute time resolution of 20 ps between points) at a bandwidth of 8 GHz. Due to the oscilloscope’s slow acquisition rate in this configuration (roughly 1000 events written to disk per minute) and limited magnet operation time, runs were taken at only four magnetic field values at the nominal HV setting. A significantly larger data set utilizing the oscilloscope and femtosecond laser was taken several months after this one and will be detailed in a subsequent publication.

The offline extraction of the timing resolutions closely follows what was done for the digitizer data. The rising edges of the trigger and signal pulses were fit with linear functions, and the time interval between the midpoints of these lines was recorded. The resulting distributions were then fit with a Gaussian function and the $\sigma$ values plotted as the timing resolution. While the superior sampling rate and bandwidth of the oscilloscope should allow one to see timing resolutions of 20 ps or better \cite{Lyashenko}, the resolution is dominated by the jitter from the PiLas laser and the trigger logic with the result that the oscilloscope results closely mirror those from the digitizers, as seen in Fig.~\ref{fig:Timing2}.

\section{Discussion} 

\subsection{Saturation due to signal rate}

\begin{figure}[ht]
\centering
\includegraphics[width=\linewidth]{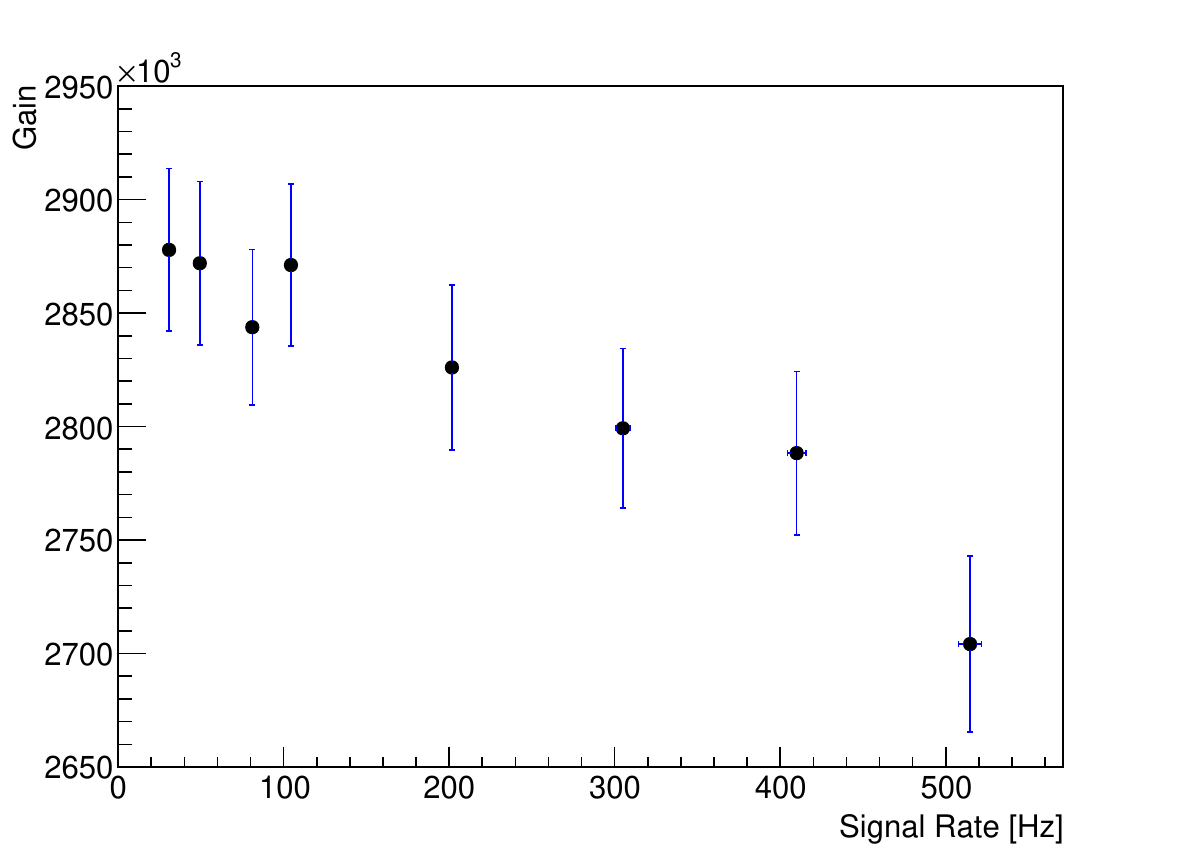}
\caption{Estimated gain as a function of signal rate. The measurements in the previous sections were performed near the third data point from left.}
\label{fig:Rate}
\end{figure}

The aforementioned measurements were conducted at a trigger rate of 1.6 kHz and with a laser pulse intensity that led to a fraction of $\sim$6\% events containing a measurable signal. To investigate systematic effects arising from the dead time of the microchannels, a signal rate scan was performed by tuning the trigger rate. The laser pulse intensity was kept constant at $\sim$15\% lower than that used in the gain measurements to further suppress the probability of multiphoton events. Here, the signal rate is defined as the product of the trigger rate and the fraction of events containing a signal above threshold. The measured gain as a function of signal rate is presented in Fig.~\ref{fig:Rate}. The observed reduction in gain at higher signal rates indicates a saturation effect on the order of several percent. Consequently, in low-rate applications, a corresponding gain increase of a few percent is expected as the microchannels operate farther from the charge-depletion regime. The laser beam has a spot diameter of only 1.5 mm, illuminating a limited number of MCP pores. Therefore, the measured signal rate primarily reflects the recovery behavior of the illuminated pores rather than that of the entire device.

\FloatBarrier

\subsection{Saturation of amplification in the second MCP}
As shown in the gain spectra (see Fig.~\ref{fig:spectra}), the right-hand tail of the distributions shrinks in the presence of a magnetic field, indicating the onset of signal saturation. This effect has been investigated thoroughly~\cite{Angle1, Angle2, Angle3, Angle4} and is attributed to the influence of the magnetic field on the photoelectron trajectories, which alters the striking angle of photoelectrons entering the microchannels. In addition, when the high voltages on the two MCPs are increased independently (Fig.~\ref{fig:APR}), the second MCP (bottom MCP) does not yield a significant gain increase compared to the first (top MCP). This further observation suggests that the amplification process in the second MCP may be approaching saturation. We speculate that the saturation arises due to the effect that the electron cloud in the transfer gap is constrained by the size of the helical trajectories, reducing the number of pores that participate in the amplification process in the second MCP; consequently, an excessive number of electrons are funneled into the same pores, which can release only a finite number of secondary electrons. This interpretation is also consistent with the relatively minor dip observed at $-13^\circ$ in Fig.~\ref{fig:gain_angleX}.

\subsection{Peak in timing at low B-field strength}
We speculate that the observed evolution of the timing resolution arises from the interplay of two competing effects. The first stems from the reduction in signal amplitude caused by the magnetic field. A smaller signal amplitude increases the uncertainty in determining the pulse midpoint, thereby degrading the timing resolution. The second effect originates from the size of the helical trajectories of the electrons in the magnetic field, which reduces the number of pores participating in the charge-multiplication process within the bottom MCP. This reduction in the number of active pores decreases transit-time fluctuations and consequently improves the timing performance.
As shown in Fig.~\ref{fig:gain_bfield}, the largest reduction in gain occurs at the first magnetic-field step, from 0 to 0.37~T, leading to a noticeable degradation in the timing resolution. As the magnetic field strength increases further, the enhanced helical motion suppresses transit-time fluctuations, resulting in an improvement in timing performance above 0.37~T. However, beyond approximately 0.7~T, the continued reduction in signal amplitude becomes the dominant effect, causing the timing resolution to deteriorate again. Further studies, including detailed simulations, are required to confirm this interpretation.

\section{Conclusions}

We have tested the performance of a representative EIC HRPPD tile as a function of three critical experimental operating parameters: the HRPPD voltage configuration, the magnetic field strength, and the orientation of the HRPPD tile in the magnetic field, and have identified stable operating points for each. Taking the HV setting of 200-675-200-675-200 as the nominal operating condition, the “EIC-HRPPD” demonstrates reliable performance in Tesla-scale magnetic fields. It achieves a gain exceeding $10^6$ in magnetic fields up to 1.8~T over most inclination angles, including those envisioned for the pfRICH at ePIC, with a low afterpulsing rate of 1.34\% and a dark-count rate of $24~\mathrm{Hz/cm^2}$ at 1.3~T. In addition, the intrinsic timing resolution of the HRPPD is expected to be better than 40~ps under the experimental conditions of ePIC, if excluding the broadening effects from the laser and trigger timing jitter. Finally, our results suggest that applying asymmetric bias voltages across the two MCPs will effectively mitigate the saturation effects we have observed, at the cost of an increase in after pulsing.



\section*{Acknowledgments}
We are grateful to David Jaffe (BNL) for his valuable comments and careful review of the manuscript. We also extend our appreciating to Bill Lenz for helping to design and assemble the experimental apparatus. Finally, we thank the support staff at the BNL SMD, including Christopher Tamargo and Raymond DeSalvo, for their dedicated technical assistance. This work is supported by Brookhaven Science Associates, LLC under Contract No. DE-SC0012704 with the U.S. Department of Energy.


\bibliography{Bibliography.bib}




\end{document}